\documentclass{revtex4}
\usepackage{amsmath,amssymb,amsfonts}
\usepackage{mathtools}
\usepackage{graphicx}
\usepackage{hyperref}

\usepackage[capitalize]{cleveref}
\usepackage{bm}
\usepackage{footmisc}
\usepackage{float}

\newcommand{\be}{\begin{equation}}
\newcommand{\ee}{\end{equation}}

\newcommand{\BigO}[1]{\ensuremath{\operatorname{O}\left(#1\right)}}

\newcommand{\LLMa}{\ensuremath{\mathcal{L}_\text{M,a}}}
\newcommand{\LLMd}{\ensuremath{\mathcal{L}_\text{M,d}}}

\newcommand{\sDA}{\ensuremath{\sigma_{\text{DA}}}}
\newcommand{\nuFSR}{\ensuremath{\nu_\text{FSR}}}
\newcommand{\Leff}{\ensuremath{L_\text{eff}}}
\newcommand{\LLeff}{\ensuremath{\mathcal{L}_\text{eff}}}
\newcommand{\svib}{\ensuremath{\sigma_\text{vib}}}

\newcommand{\Dza}{\ensuremath{\Delta z_a}}
\newcommand{\wOa}{\ensuremath{w_{0,a}}}
\newcommand{\wOd}{\ensuremath{w_{0,d}}}

\begin{document}
\title{Optimal design of diamond-air microcavities for quantum networks using an analytical approach}

\author{S.B. van Dam} 
\affiliation{QuTech, Delft University of Technology, 2628 CJ Delft, The Netherlands}
\affiliation{Kavli Institute of Nanoscience, Delft University of Technology, 2628 CJ Delft, The Netherlands}

\author{M. Ruf}
\affiliation{QuTech, Delft University of Technology, 2628 CJ Delft, The Netherlands}
\affiliation{Kavli Institute of Nanoscience, Delft University of Technology, 2628 CJ Delft, The Netherlands}

\author{R. Hanson}
\affiliation{QuTech, Delft University of Technology, 2628 CJ Delft, The Netherlands}
\affiliation{Kavli Institute of Nanoscience, Delft University of Technology, 2628 CJ Delft, The Netherlands}

\begin{abstract}
Defect centres in diamond are promising building blocks for quantum networks thanks to a long-lived spin state and bright spin-photon interface.
However, their low fraction of emission into a desired optical mode limits the entangling success probability.
The key to overcoming this is through Purcell enhancement of the emission.
Open Fabry-Perot cavities with an embedded diamond membrane allow for such enhancement while retaining good emitter properties. 
To guide the focus for design improvements it is essential to understand the influence of different types of losses and geometry choices.
In particular, in the design of these cavities a high Purcell factor has to be weighed against cavity stability and efficient outcoupling. 
To be able to make these trade-offs we develop analytic descriptions of such hybrid diamond-and-air cavities as an extension to previous numeric methods. 
The insights provided by this analysis yield an effective tool to find the optimal design parameters for a diamond-air cavity.
\end{abstract}
\maketitle

\section{Introduction}
Quantum networks rely on entanglement distributed among distant nodes \cite{Reiserer2014}. Nitrogen-vacancy (NV) defect centers in diamond can be used as building blocks for such networks, with a coherent spin-photon interface that enables the generation of heralded distant entanglement \cite{Bernien2013,Gao2015}. The long-lived electron spin and nearby nuclear spins provide quantum memories that are crucial for extending entanglement to multiple nodes and longer distances \cite{Kolkowitz2012,Taminiau2012,Zhao2012,Kalb2017,Humphreys2018}. However, to fully exploit the NV centre as a quantum network building block requires increasing the entanglement success probability. One limitation to this probability is the low efficiency of the NV spin-photon interface. Specifically, entanglement protocols depend on coherent photons emitted into the zero-phonon line (ZPL), which is only around $3\%$ of the total emission \cite{Riedel2017a}, and collection efficiencies are finite due to limited outcoupling efficiency out of the high-refractive index diamond. These can both be improved by embedding the NV centre in an optical microcavity at cryogenic temperatures, benefiting from Purcell enhancement \cite{Purcell1946, Faraon2011, Barclay2011,Faraon2012,Hausmann2013,Li2015,Riedrich-Moller2015,Johnson2015}. A promising cavity design for applications in quantum networks is an open Fabry-Perot microcavity with an embedded diamond membrane \cite{Hunger2010,Janitz2015,Bogdanovic2017,Riedel2017a}. Such a design provides spatial and spectral tunability and achieves a strong mode confinement while the NV centre can reside in the diamond membrane far away ($\approx\mu$m) from the surface to maintain bulk-like optical properties. 
 
The overall purpose of the cavity system is to maximise the probability to detect a ZPL photon after a resonant excitation pulse.
This figure of merit includes both efficient emission into the ZPL into the cavity mode, and efficient outcoupling out of the cavity.
The core requirement is accordingly to resonantly enhance the emission rate into the ZPL. However this must be accompanied by vibrational stability of the system; an open cavity design is especially sensitive to mechanical vibrations that change the cavity length, bringing the cavity off-resonance with the NV centre optical transition. Furthermore the design should be such that the photons in the cavity mode are efficiently collected.
We aim to optimize the cavity parameters in the face of these (often contradicting) requirements. For this task, analytic expressions allow the influence of individual parameters to be clearly identified and their interplay to be better understood. In this manuscript we take the numerical methods developed in \cite{Janitz2015} as a starting point, and find the underlying analytic descriptions of hybrid diamond-air cavities. We use these new analytic descriptions to investigate the optimal parameters for a realistic cavity design.

We define two boundary conditions for the design of the cavity, within which we operate to maximize the figure of merit: the probability to detect a ZPL photon. 
Firstly we require the optical transition to be little influenced by decoherence and spectral diffusion so that the emitted photons can be used for generating entanglement between remote spins \cite{Bernien2013}. Showing enhancement of the ZPL of narrow linewidth NV centres is still an outstanding challenge. The demonstration in \cite{Riedel2017a} employed NV centres in a 1 $\mu$m thick membrane with optical transitions with a linewidth under the influence of spectral diffusion of $\approx1$ GHz, significantly broadened compared to the $\approx13$ MHz lifetime-limited value. While the mechanism of broadening is not fully understood, using a thicker diamond may be desired. We therefore conservatively use a diamond membrane thickness of 4 $\mu$m in the simulations throughout this manuscript.
Secondly, we consider a design that enables long uninterrupted measurements at cryogenic temperatures, potentially at a remote location with no easy access (such as a data center), through the use of a closed-cycle cryostat. The vibrations induced by the cryostat's pulse tube can be largely mitigated passively \cite{Bogdanovic2017}. Active stabilisation of Fabry-Perot cavities has been demonstrated \cite{Gallego2015,Brachmann2016}, including at a high bandwidth \cite{Khudaverdyan2008,Janitz2017}, however these results have not yet been extended to operation in a pulse-tube cryostat. In the simulations in this manuscript we therefore assume that vibrations lead to passively stabilised cavity length deviations of 0.1 nm RMS \cite{Bogdanovic2017}.
While these boundary conditions influence the simulated maximally achievable probability to detect a ZPL photon, the analytic descriptions in this manuscript are not limited to these parameter regimes.

The layout of this manuscript is the following. We start by describing the one-dimensional properties of the cavities in \cref{sec:longitudinalconfinement}. These are determined by the distribution of the electric field over the diamond and air parts of the cavity and its impact on the losses out of the cavity. In \cref{sec:transversalconfinement} we extend this treatment to the transverse extent of the cavity mode, analysing the influence of the geometrical parameters. Finally we include real-world influences of vibrations and unwanted losses to determine the optimal mirror transmittivity and resulting emission into the ZPL in \cref{sec:optimal_finesse}.

\section{The one-dimensional structure of a hybrid cavity} \label{sec:longitudinalconfinement}
The resonant enhancement of the emission rate in the ZPL is determined by the Purcell factor \cite{Purcell1946,Fox2006}:
	\begin{equation} \label{eq:Fp}
	F_p = \xi \frac{3 c \lambda_0^2}{4 \pi n_d^3}\frac{1}{\delta \nu V},
	\end{equation}
	where $\xi$ describes the spatial and angular overlap between the NV centre's optical transition dipole and the electric field in the cavity; $c$ is the speed of light, $\lambda_0$ is the free-space resonant wavelength and $n_d$ the refractive index in diamond. $\delta \nu$ is the cavity linewidth (full width at half maximum (FWHM) of the resonance that we assume to be Lorentzian), and $V$ is the mode volume of the cavity.
	While the ZPL emission rates can be enhanced through the Purcell effect, the off-resonant emission into the phonon side band (PSB) will be nearly unaffected in the parameter regimes considered. This is the result of the broad PSB transition linewidth ($\delta\nu_{PSB}$ is several tens of THz), that leads to a reduced effective quality factor, replacing $\nu/\delta\nu\rightarrow \nu/\delta\nu + \nu/\delta\nu_{PSB}$ \cite{Gerard2003}. This results in a low Purcell factor for the PSB. Selection rules for the optical transitions further prevent enhancement of the ZPL emission rate to ground states other than the desired one. 
		The resulting branching ratio of photons into the ZPL, into the cavity mode is therefore \cite{Faraon2011,Li2015}:
	\begin{equation} \label{eq:branching}
	\beta = \frac{\beta_0 F_p}{\beta_0 F_p +1},
	\end{equation}
	where $\beta_0$ is the branching ratio into the ZPL in the absence of the cavity. Values for $\beta_0$ have been found in a range $\approx 2.4-5\%$ \cite{Faraon2011,Riedel2017a}; we here use $\beta_0=3\%$. Note that to maximize the branching ratio we should maximize the Purcell factor, but that if $\beta_0 F_p\gg1$ the gain from increasing $F_p$ is small.

To optimize the Purcell factor through the cavity design we should consider the cavity linewidth and mode volume. In this section we focus on the linewidth of the cavity, that is determined by the confinement of the light between the mirrors. In \cref{sec:transversalconfinement} we evaluate the mode volume of the cavity.

The cavity linewidth is given by the leak rate out of the cavity: $\delta \nu = \kappa /( 2 \pi) $. For a general bare cavity this can be expressed as:
\begin{equation} \label{eq:dnu_bare}
\delta\nu = \frac{1}{2\pi} \frac{\text{losses per round-trip}}{\text{round-trip duration}}=\frac{1}{2\pi}\frac{ \mathcal{L}}{2nL/c}=\frac{c/(2nL)}{2\pi/\mathcal{L}} = \nuFSR/F,
\end{equation}
for a cavity of length $L$ in a medium with refractive index $n$. $\mathcal{L}$ are the losses per round-trip. In the last two steps we have written the expression such that one can recognize the standard definitions of free spectral range ($\nuFSR=c/(2nL)$) and Finesse ($F=2\pi/\mathcal{L}$). By using this description we assume that the losses per round trip are independent of the cavity length, which is true if losses appear at surfaces only.

\begin{figure}[t]%
\centering
\includegraphics[width=0.7\columnwidth]{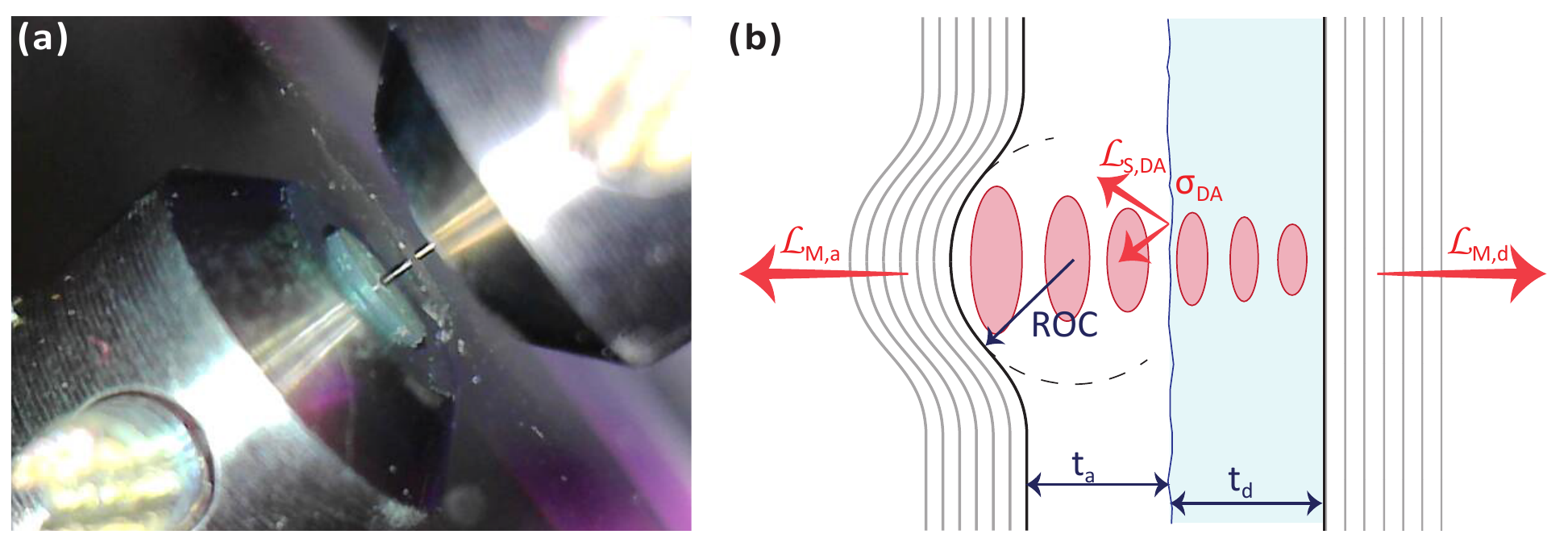}%
\caption{{\bf Plane-concave fiber-based microcavities.} (a) Experimental plane-concave fiber-based microcavity. The cavity is formed at the fiber tip. Reflections of the fiber and holders are visible in the mirror. (b) The geometry of an open diamond-air cavity is described by the diamond thickness $t_d$, air gap $t_a$, and the dimple radius of curvature (ROC). The most important losses are through the mirror on the air-side ($\LLMa$) and on the diamond-side ($\LLMd$), and from scattering on the diamond-air interface ($\mathcal{L}_\text{S,DA}$) resulting from a rough diamond surface with surface roughness $\sDA$.}
\label{fig:cavity_schematic}%
\end{figure}

For a hybrid diamond-air cavity (\cref{fig:cavity_schematic}) this definition does not work anymore: due to the partially reflective interface between diamond and air, we cannot use the simple picture of a photon bouncing back and forth in a cavity. Instead, we should consider the electric field mode and its relative energy density in each part of the cavity. Staying close to the formulations used for a bare cavity, and choosing the speed of light in the diamond part ($c/n_d$) as a reference, the duration of an \emph{effective round-trip} is $c/(2n_d\Leff)$, where $\Leff$ is an effective cavity length. This effective length should contain the diamond thickness and the width of the air gap weighted by the local energy density of the photon mode, relative to the energy density in the diamond membrane. Generalising this, the effective length of the cavity system can be described by the `energy distribution length' \cite{Greuter2014}:   
\begin{equation} \label{eq:L-eff} 
L_{\text{eff}} \equiv \frac{\int_{cav} \epsilon(z) |E(z)|^2 \mathrm{d} z}{\epsilon_0 n_d^2 |E_{max,d}|^2/2}.
\end{equation} 
In this formulation $\epsilon=\epsilon_0 n^2$ is the permittivity of a medium with refractive index $n$, $E(z)$ is the electric field in the cavity and $E_{max,d}$ is the maximum electric field in diamond. The integral extends over the full cavity system, such that the effective length automatically includes the penetration depth into the mirrors.
The resulting formulation for the linewidth of a hybrid cavity analogous to \cref{eq:dnu_bare} is: 
	\begin{equation} \label{eq:dnu}
	\delta\nu = \frac{c/(2 n_d \Leff)}{(2\pi/\LLeff)},
	\end{equation}
where $\LLeff$ are the losses encountered during the effective round-trip. Here, like in the bare cavity case, we assume these losses to occur only at surfaces. This is a realistic assumption since the most important losses are expected to be from mirror transmission and absorption and diamond surface roughness.

In the above we have taken the field in diamond as reference for the effective round-trip. This choice is motivated by the definition of the mode volume as the integral over the electric field in the cavity relative to the electric field at the position of the NV centre - in diamond. It is given by \cite{Gerard2003,Sauvan2013}:
 \begin{equation} \label{eq:mode_volume-1}
V= \frac{ \int_{cav} \epsilon(\vec{r}) |E(\vec{r})|^2  \mathrm{d}^3\vec{r}}{\epsilon(\vec{r_0}) |E(\vec{r}_0)|^2},
\end{equation}
with $\vec{r}_0$ the position of the NV centre, that we assume to be optimally positioned in an antinode of the cavity field in diamond, such that $E(\vec{r}_0) = E_{max,d}$. We choose to explicitly include effects from sub-optimal positioning in the factor $\xi$ in the Purcell factor (\cref{eq:Fp}) rather than including them here. If we evaluate the integral in the radial direction we see that the remaining integral describes the effective length as defined above: 
\begin{equation}\label{eq:mode_volume-2}
V =  \frac{\pi w_0^2}{2}  \frac{ \int_{cav} \epsilon(z) |E(z)|^2  \mathrm{d}z}{\epsilon_0 n_d^2 |E_{max,d}|^2} = \frac{\pi w_0^2}{4} \Leff,
\end{equation}
where $w_0$ is the beam width describing the transverse extent of the cavity mode at the NV, that we will come back to in \cref{sec:transversalconfinement}.
We notice that the effective length appears in both the linewidth and the mode volume. In the Purcell factor ($F_p\sim 1/(\delta\nu V)$), the effective length cancels out. This is the result of our assumption that the losses per round-trip occur only at surfaces in the cavity. 

The parameter relevant for Purcell enhancement in \cref{eq:dnu} is thus $\LLeff$. Since these are the losses in an effective round-trip, we expect that they depend on the electric field distribution. We therefore first analyse the electric field distribution in the following section, before finding the effective losses related to the mirror losses and diamond surface scattering in \cref{sec:mirrorlosses,sec:scattering}.

\subsection{Electric field distribution over diamond and air}\label{sec:fielddistribution}
The electric field distribution in the cavity on resonance is dictated by the influence of the partially reflective diamond-air interface. If the two parts were separated, the resonant mode in air would have an antinode at this interface, but the mode in diamond would have a node at that position. These cannot be satisfied at the same time, such that in the total diamond-air cavity system the modes hybridize, satisfying a coupled system resonance condition \cite{Janitz2015,Bogdanovic2017,suppl}. Two special cases can be distinguished for these resonant modes: the `air-like mode', in which the hybridized mode has an antinode at the diamond-air interface, and the `diamond-like mode' in which there is a node at the interface. For a fixed resonance frequency matching the NV-centre's ZPL emission frequency ($\approx 470.4$ THz), the type of mode that the cavity supports is fully determined by the diamond thickness. The tunable air gap allows for tuning the cavity to satisfy the resonance condition for any frequency. 

\begin{figure}[H] %
\centering
\includegraphics[width=0.9\columnwidth]{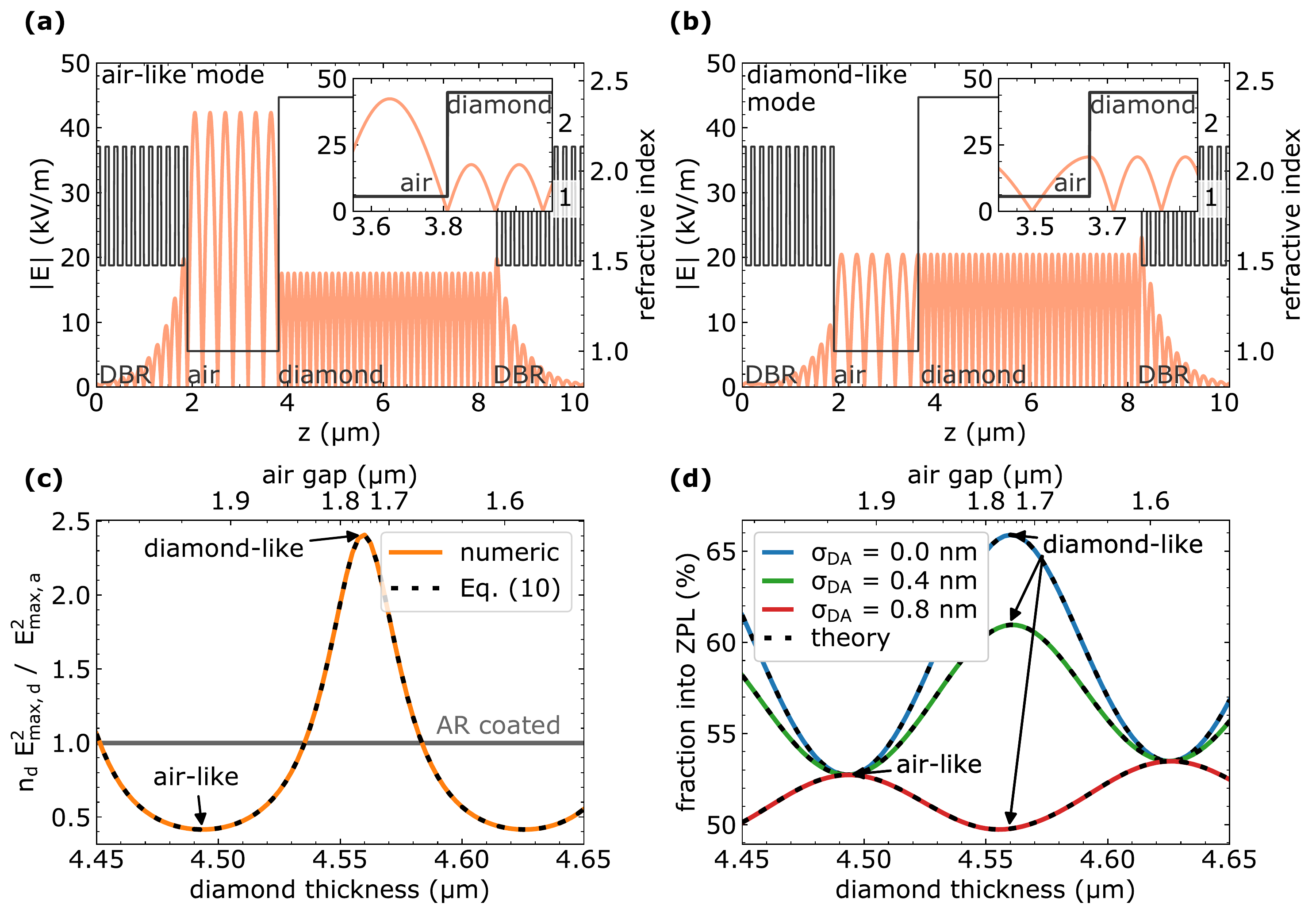}
\caption{{\bf Diamond-like and air-like modes in a diamond-based microcavity.} (a,b) The electric field strength (orange, left axis) in a diamond-air cavity satisfying the conditions for (a) an air-like mode and (b) a diamond-like mode is calculated using a transfer matrix model. (c) The relative intensity of light in the diamond membrane and air gap is described by \cref{eq:rel_intensity}. It oscillates between $n_d\approx2.41$ for the diamond-like mode and $1/n_d\approx0.4$ for the air-like mode. When the diamond is anti-reflection (AR) coated, the oscillations vanish. To stay on the same resonance for varying diamond thickness the air gap is tuned. The corresponding values on the top x-axis do not apply to the cavity with AR coating. (d) The fraction of photons emitted into the ZPL shows a strong dependency on the diamond thickness, presented for three values of RMS diamond roughness $\sigma_{DA}$. The emission into the ZPL is determined from \cref{eq:Fp,eq:branching}, with the mode volume as described in \cref{sec:transversalconfinement}. The linewidth is numerically found from the transfer matrix model (solid lines) or with analytic descriptions using \cref{eq:dnu} together with \cref{eq:L_M,eq:L_SAD} (black dashed lines). The mirror transmittivity corresponds to a distributed Bragg reflector (DBR) stack with 21 alternating layers of Ta$_2$O$_5$ ($n=2.14$) and SiO$_2$ ($n=1.48$) (giving $\mathcal{L}_{M,A}=260$ ppm and $\mathcal{L}_{M,D} = 630$ ppm). The dimple radius of curvature used is $ROC=25~\mu$m.} %
\label{fig:diamond_air_modes}
\end{figure}

Using a transfer matrix model \cite{Orfanidis2016,Janitz2015} we find the electric field distribution for both the air-like and the diamond-like modes, as shown in \cref{fig:diamond_air_modes}(a) and (b). 
If the cavity supports a diamond-like mode, the field intensity (proportional to $n E_{max}^2 $\cite{Dodd1991}) is higher in the diamond-part, and vice-versa for the air-like mode. 
The relative intensity of the electric field in the cavity in the diamond membrane compared to the air gap is shown in \cref{fig:diamond_air_modes}(c) for varying diamond thicknesses. The relation that the relative intensity satisfies can be explicitly inferred from the continuity condition of the electric field at the diamond-air interface:
	\begin{equation}  \label{eq:E_continuitity} 
	E_{max,a}\sin(\frac{2\pi t_a}{\lambda_0}) = E_{max,d}\sin(\frac{2\pi t_dn_d}{\lambda_0});
	\end{equation}
where the air gap $t_a$ corresponds to the hybridized diamond-air resonance condition \cite{suppl}: 
 \begin{equation}
t_a = \frac{\lambda_0}{2\pi}\arctan\left(-\frac{1}{n_d} \tan\left(\frac{2\pi n_dt_d}{\lambda_0}\right)\right)+\frac{m\lambda_0}{2},
\end{equation}
for an integer $m$. We use $n_{air}=1$.
The relative intensity in the air gap can thus be written as 
\begin{equation} \label{eq:rel_intensity}
\frac{E_{max,a}^2}{n_d E_{max,d}^2} = \frac{1}{n_d}\sin^2\left(\frac{2\pi n_dt_d}{\lambda_0}\right)+n_d\cos^2\left(\frac{2\pi n_dt_d}{\lambda_0}\right).
\end{equation}
This ratio reaches its maximal value $n_d$ for an air-like mode, while the minimal value $1/n_d$ is obtained for a diamond-like mode. This relation is shown in \cref{fig:diamond_air_modes}(c) as a dashed line, that overlaps with the numerically obtained result.

To remove the mixing of diamond-like and air-like modes, an anti-reflection (AR) coating can be applied on the diamond surface. This is in the ideal case a layer of refractive index $n_{AR}=\sqrt{n_d} \approx 1.55$ and thickness $t_{AR}=\lambda_0/(4n_{AR})$. The effect of a coating with refractive index $n_{AR}$ is shown as a gray line in \cref{fig:diamond_air_modes}(c). For a realistic coating with a refractive index that deviates from the ideal, a small diamond thickness-dependency remains \cite{suppl}.  

Next we determine the diamond thickness-dependency of an NV centre's branching ratio into the ZPL \cite{code}. For this we need to find the linewidth and mode volume: we use the transfer matrix to numerically find the cavity linewidth from the cavity reflectivity as a function of frequency, and we calculate the mode volume using \cref{eq:mode_volume-2}. The method with which we determine the beam waist $w_0$ will be later outlined in \cref{sec:transversalconfinement}. We further assume that the NV center is optimally placed in the cavity. To include the effect of surface roughness we extend the Fresnel reflection and transmission coefficients in the matrix model as described in \cite{Filinski1972,Szczyrbowski1977,Katsidis2002,Janitz2015}\cite{suppl}.
\Cref{fig:diamond_air_modes}(d) shows that the resulting emission into the ZPL is strongly dependent on the electric field distribution over the cavity, both for the cases with and without roughness of the diamond interface. 

Since we have already seen that the effective cavity length does not appear in the final Purcell factor, the varying emission into the ZPL with diamond thickness has to originate from varying effective losses in \cref{eq:dnu}. In the next paragraphs we develop analytic expressions for the effective losses that indeed exhibit this dependency on the electric field distribution. We address the two most important sources of losses in our cavity: mirror losses and roughness of the diamond-air interface.

\subsection{Mirror losses} \label{sec:mirrorlosses}
As described at the start of this section the mirrors on either side of a bare cavity are encountered once per round-trip, making the total mirror losses simply the sum of the individual mirror losses. For a hybrid cavity, we have rephrased the definition of linewidth to \cref{eq:dnu} by introducing an effective round-trip. In this picture, the mirrors on the diamond side are encountered once per round-trip, while the losses on the air side should be weighted by the relative field intensity in the air part. The resulting effective mirror losses 
are described by:
\begin{equation} \label{eq:L_M}
\mathcal{L}_{M,\text{eff}} = \frac{E_{max,a}^2}{n_d E_{max,d}^2}  \mathcal{L} _{M,a} + \mathcal{L} _{M,d}, 
\end{equation}
where $\mathcal{L} _{M,a}$ are the losses of the mirror on the air side, $\mathcal{L} _{M,d}$ the losses of the diamond side mirror and the relative intensity in the air gap is given by \cref{eq:rel_intensity}.  
Since this factor fluctuates between $1/n_d$ for the diamond-like mode and $n_d$ for the air-like mode, the effective losses are lower in the diamond-like mode than in the air-like mode. 
This results in the strong mode-dependency of the emission into the ZPL in \cref{fig:diamond_air_modes}(d). 
The analytic expresstion for the effective mirror losses can be used to calculate the fraction of NV emission into the ZPL, resulting in the black dashed line in \cref{fig:diamond_air_modes}(d). This line overlaps with the numerically obtained result. Our model using the effective round-trip thus proves to be a suitable description of the system.

In \cref{fig:effective_losses}(a) the effective losses are plotted for a relative contribution of $\LLMa$ to the total mirror losses, that are fixed. If this contribution is larger, the deviations between the effective mirror losses in the diamond-like and air-like mode are stronger.

For a cavity with an AR coating ($E_{max,a}^2 = n_d E_{max,d}^2$) the losses would reduce to the standard case $\mathcal{L}_{M,a}+ \mathcal{L} _{M,d}$ as expected. From the perspective of fixed mirror losses the best cavity performance can thus be achieved in a cavity without AR coating, supporting a diamond-like mode.

\subsection{Scattering at the diamond-air interface}  \label{sec:scattering}
Next to mirror losses the main losses in this system are from scattering due to diamond roughness. The strength of this effect depends on the electric field intensity at the position of the interface. 

The electric field intensity at the diamond-mirror interface depends on the termination of the distributed Bragg reflector (DBR). If the last DBR layer has a high index of refraction, the cavity field has an node at this interface, while if the refractive index is low the field would have a antinode there. The losses due to diamond surface roughness are thus negligible with a high index of refracted mirror. Such a mirror is therefore advantageous in a cavity design, even though a low index of refraction termination interfaced with diamond provides lower transmission in a DBR stack with the same number of layers \cite{Janitz2015}. We assume a high index of refraction mirror termination and thus negligible surface roughness losses throughout this manuscript. The mirror transmissions specified already take the interfacing with diamond into account. 

At the diamond-air interface the field intensity depends on the type of the cavity mode. The air-like mode (with a node at the interface) is unaffected, while the diamond-like mode is strongly influenced (\cref{fig:diamond_air_modes}(d) and \cref{fig:effective_losses}(a), green and red lines). 

From a matching matrix describing a partially reflective rough interface \cite{Filinski1972,Szczyrbowski1977,Katsidis2002,Janitz2015}, we can find the effective losses at the interface. To get the effective losses on one side of the interface, we find the difference between the intensity of the field travelling towards the surface and the intensity of the field travelling back. The field travelling away from the interface contains contributions both from the reflected field, as well as from the field transmitted through the other side of the interface. For one side, this is thus described as:
\begin{align}
\mathcal{L}_{S,12} & = 1- |E_{1,out}|^2/|E_{1,in}|^2\\
& = 1- |\rho'_{12} E_{1,in} + \tau'_{21} E_{2,in}|^2/|E_{1,in}|^2,
\end{align}
where $E_{1,in}$ and $E_{2,in}$ are the incoming field from the left-hand side and right-hand side of the interface respectively. $E_{1,out}$ is the outgoing field on the left-hand side of the interface. Furthermore, $\rho'_{12}$ and $\tau'_{21}$ are the reflection and transmission coefficients extended to include surface roughness.

We evaluate this expression for losses from the diamond-side and from the air-side, multiplying the latter by the relative intensity (\cref{eq:rel_intensity}) as we did in the case for the mirror losses. The resulting losses per effective round-trip are:
\begin{align}
\mathcal{L}_{S,\text{eff}} &= \mathcal{L}_{S,DA} + \frac{E_{max,a}^2}{n_d E_{max,d}^2} \mathcal{L}_{S,AD}\\ 
& \approx \sin^2\left(\frac{2\pi n_d t_d}{\lambda_0}\right) \frac{(1+n_d)}{n_d} \left(1-n_d\right)^2\left(\frac{4\pi\sigma_{DA}}{\lambda_0}\right)^2.\label{eq:L_SAD}
\end{align}
In the evaluation of this expression we use a Taylor series approximation for the exponents in the reflection and transmission coefficients, and keep terms up to $\BigO{(4\pi\sigma_{DA}/\lambda_0)^2}$. A detailed derivation can be found in the Supplementary Information \cite{suppl}.
This description matches well with the numerically found result, which is evidenced in \cref{fig:diamond_air_modes}(d) where the gray dashed lines obtained with \cref{eq:L_SAD} overlap with the numerical description (green and red lines). 

In the case that the diamond would be AR coated, the coating roughness is expected to follow the diamond roughness. In this case, scattering losses are always present, with only a small modification based on the exact diamond thickness. The amount of scattering losses is however lower than in the diamond-like mode.

	\subsection{Minimizing the effective losses}  \label{sec:diamond-air-tradeoff}
Assuming that mirror losses and scattering at the air-diamond interface are the main contributors to the losses, the total effective losses are $\LLeff = \mathcal{L}_{M,\text{eff}} + \mathcal{L}_{S,\text{eff}}$. Other losses could originate from absorption in the diamond or clipping losses (see \cref{sec:clippinglosses}), but have a relatively small contribution in the considered parameter regimes \cite{Janitz2015}.

As described above, an AR coating on the diamond membrane ensures that the intensities of the electric field in diamond and air are the same, while they would otherwise fluctuate with the diamond thickness. The mirror losses are then independent of diamond thickness, and the scattering losses are close to constant. The mirror losses with an AR coating are higher than the losses in the case of no AR coating in the diamond-like mode for the same mirror parameters. The scattering losses however are lower with an AR coating than in the diamond-like mode. Whether the highest Purcell factor can be achieved with or without AR coating thus depends on the relative losses. For the parameters in \cref{fig:diamond_air_modes} if the roughness is $<$ 0.4 nm a higher Purcell factor can be achieved in the diamond-like mode without an AR coating than with an AR coating.

If the diamond is not AR coated, we can decide to select either a diamond-like or air-like mode. From the previous section we see that $\mathcal{L}_{M,\text{eff}}$ is lowest for the diamond-like mode, while $\mathcal{L}_{S,\text{eff}}$ is largest in that case. Whether a system supporting an air-like or a diamond-like mode is preferential depends on their relative strength. To be able to pick this freely requires tuning of the diamond thickness on the scale $\lambda_0/(4 n_d)=66$ nm, or using the thickness gradient of a diamond membrane to select the regions with the preferred diamond thickness. Note that the diamond thickness does not have to be tuned exactly to the thickness corresponding to a diamond-like mode. From \cref{fig:diamond_air_modes}(c) it is clear that the effective mirror losses are reduced compared to the AR coating value in a thickness range of $\approx 40$ nm around the ideal diamond-like value, corresponding to about 35\% of all possible diamond thicknesses. 

\begin{figure}[H] 
\centering
\includegraphics[width=0.9\columnwidth]{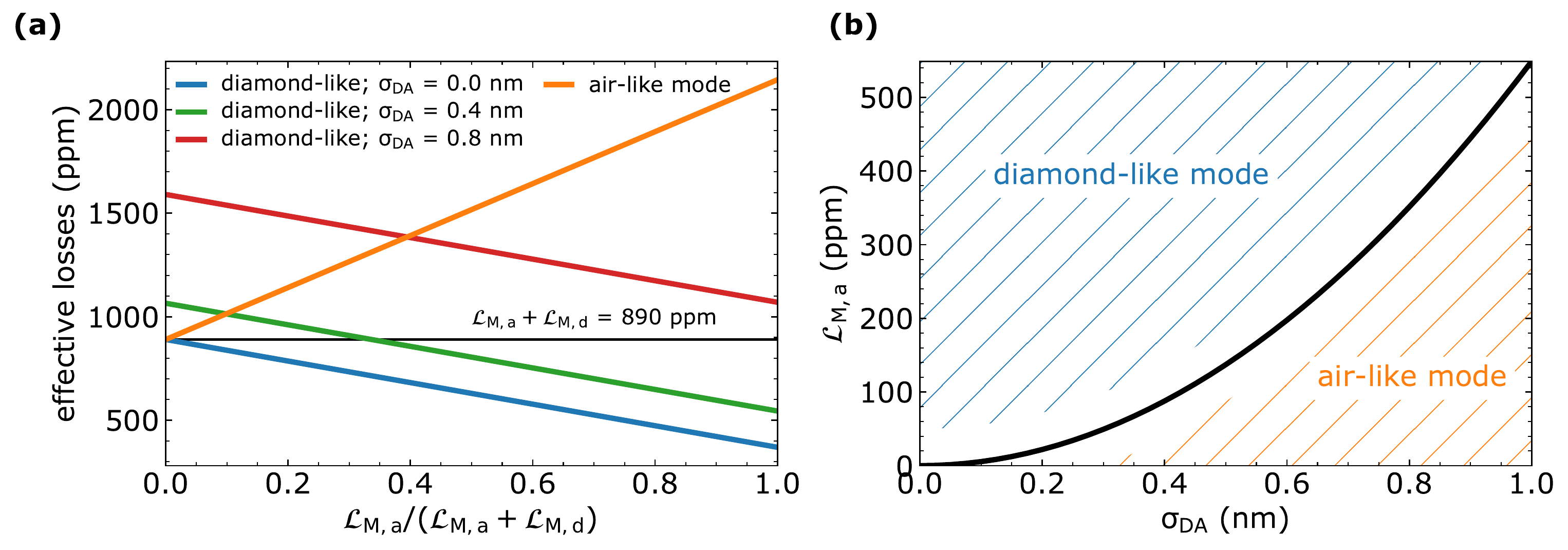}%
\caption{{\bf Effective losses in a diamond-based microcavity.} (a) The effective losses in the cavity depend on whether the cavity supports a diamond-like or air-like mode. The difference is most pronounced if the losses on the air side are dominant. For the fixed value of $\LLMa+\LLMd=890$ ppm shown, the effective losses can be up to $\approx 2150$ ppm in the air-like mode (orange line), or as low as $470$ ppm in the diamond-like mode (blue line). Scattering on the diamond-air interface (green and red lines) increase the losses in the diamond-like mode, but do not affect the air-like mode. (b) Depending on the bare losses on the air mirror and the amount of diamond surface roughness the total losses are lowest in the diamond-like mode (shaded region above the black curve) or the air-like mode (below the curve).} %
\label{fig:effective_losses}
\end{figure}

Using the analytic expressions for the losses (\cref{eq:L_M,eq:L_SAD}) we can decide whether being in a diamond-like and air-like is beneficial. If the total losses in the diamond-like mode are less than the total losses in the air-like mode, it is beneficial to have a cavity that supports a diamond-like mode. This is the case if: 
	\begin{equation} \label{eq:DA_tradeoff}
	\left(\frac{4\pi\sigma_{DA}}{\lambda_0}\right)^2 \frac{(n_d+1)(n_d-1)^2}{n_d} < \left( n_d  - \frac{1}{n_d} \right)\LLMa.  \\\
	\end{equation}
	\Cref{fig:effective_losses}(b) shows the \LLMa~for varying $\sigma_{\text{DA}}$ for which both sides of the above expression are equal. In the region above the curve, where \cref{eq:DA_tradeoff} holds, the best Purcell factor is achieved in the diamond-like mode. In the region below the curve, the Purcell factor is maximized for the air-like mode.
	\\
	
	Concluding, to achieve the highest Purcell factor low losses are key. These losses are strongly influenced by whether the cavity supports diamond-like or air-like modes. Analytic descriptions of the mirror losses and losses from diamond surface roughness depending on the electric field distribution, enable to find whether a diamond-like or air-like mode performs better.

\section{Transverse extent of Gaussian beams in a hybrid cavity}  \label{sec:transversalconfinement}
Having analysed the one-dimensional structure of the cavity, we turn to the transverse electric field confinement. We have seen in \cref{eq:mode_volume-2} that the mode volume can be described as
	\begin{equation} \label{eq:V_mod}
	V = \frac{\pi w_{0,d}^2}{4} L_{\text{eff}} \equiv g_0\left(\lambda_0/n_d\right)^2 L_{\text{eff}},
	\end{equation}
where we define a geometrical factor $g_0 \equiv \frac{\pi w_{0,d}^2}{4}/(\frac{\lambda_0}{n_d})^2$, and $w_{0,d}$ is the beam waist in diamond. Since $L_{\text{eff}}$ cancels out in the Purcell factor, $g_0$ captures all relevant geometrical factors in the mode volume. Note that combining \cref{eq:Fp} with \cref{eq:V_mod,eq:dnu}, the Purcell factor can be written as $F_p=3 \xi /(g_0\mathcal{L}_{\text{eff}})$.

In this section we describe how to find the beam waist $w_{0,d}$, and which parameters play a role in minimizing it. Furthermore, we quantify the losses resulting if the beam extends outside of the dimple diameter.

	\subsection{Beam waist} \label{sec:beamwaist}
We describe the light field in our cavity using a coupled Gaussian beams model \cite{Janitz2015}. The hybrid cavity supports two Gaussian beams: one that lives in the air gap of the cavity, and one in the diamond (\cref{fig:dimple_parameters}(a), indicated in orange and blue respectively). The boundary conditions for the model are provided by the diamond thickness, width of the air gap and the radius of curvature (ROC) of the fiber dimple \cite{suppl}. In the model we assume that the diamond surface is planar. We note that this deviates from the assumption in \cite{Janitz2015}, where the diamond surface is assumed to follow the beam curvature at the interface. The latter assumption would introduce a lensing effect, leading to a narrower effective beam waist than for a plane surface. The planar interface causes mixing with higher-order modes, but the influence of these effects is expected to be small due to the large radius of curvature of the mode at the interface \cite{Janitz2015}.

A solution to this model provides the beam waist of both beams ($\wOd$ and $\wOa$) and the related Rayleigh lengths ($z_{0,d}$, $z_{0,a}$) as well as the location of the beam waist of the air beam with respect to the plane mirror, $\Dza$. Previously such a model has been solved numerically \cite{Janitz2015}, but an analytic solution gives insight in the influence of the individual cavity parameters. The analytic solution that we find is given by \cite{suppl}:
	\begin{align}
w_{0,a}&= w_{0,d}, ~~~~(\rightarrow z_{0,a} \approx z_{0,d}/n_d); \label{eq:w_0a_vs_w_0d} \\
\Dza &= t_d \left(1-\frac{1}{n_d}\right); \label{eq:delta_za}\\
w_{0,d}&= \sqrt{\frac{\lambda_0}{\pi}}  \left(\left(t_a + \frac{t_d}{n_d}\right)\left(ROC-\left(t_a+\frac{t_d}{n_d} \right) \right)\right)^{1/4}. \label{eq:w_0a}
\end{align}

In the last expression for the beam waist we recognize the standard expression for the beam waist of a plane-concave cavity \cite{Hunger2010}, but with a new term taking the position of cavity length: 
\begin{equation} \label{eq:L_mod}
L' \equiv  t_a +\frac{t_d}{n_d} ~(= t_a + t_d - \Dza).
\end{equation}
As an important result, the influence of the diamond thickness is a factor $1/n_d \approx 0.42$ less than that of the width of the air gap. We indeed see in \cref{fig:dimple_parameters}(c) and (d) that increasing the air gap from 1 to 4 $\mu$m (green line) has a larger effect on $\wOd$ and $g_0$ than increasing the diamond thickness from 1 to 4 $\mu$m (orange line). 

The minimal air gap that can be achieved is set by the dimple geometry (see \cref{fig:dimple_parameters}(b)). Smooth dimples with a small ROC can be created in several ways, including with CO$_2$ laser ablation or focused-ion-beam milling of optical fibers or fused silica plates \cite{Dolan2010,Hunger2010,Barbour2011,Trichet2015,Najer2017}. The dimple depth for dimple parameters as considered here is typically $z_d \approx 0.2-0.5~\mu$m, while a relative tilt between the mirror of an angle $\theta$ introduces an extra distance of $z_f = D_f/2 \sin(\theta)\approx D_f\theta/2$, which is $\approx 4~\mu$m for a fiber cavity \cite{suppl}.
 This last effect if thus dominant over the dimple depth. To reduce the minimal air gap in fiber-based cavities, the most important approach to lowering the mode volume is thus by shaping the fiber tip \cite{Kaupp2016}. 
For cavities employing silica plates the large extent of the plates demands careful parallel mounting of the mirror substrates.

\subsection{Clipping losses} \label{sec:clippinglosses}
\begin{figure}[H]%
	\centering
\includegraphics[width=0.8\textwidth]{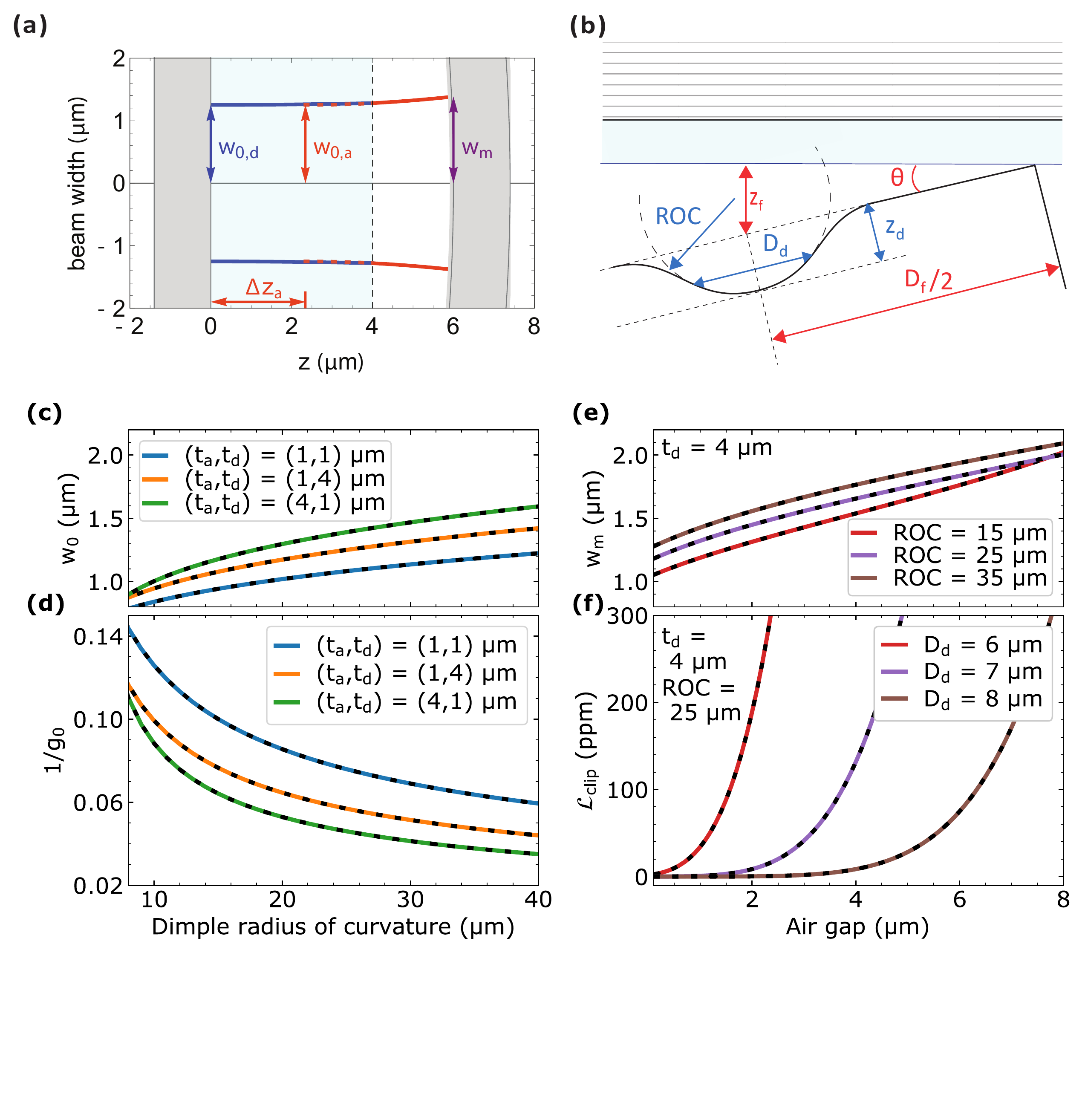}%
\caption{{\bf Transverse extent of Gaussian beams in a microcavity.} (a) The transverse extent of the cavity mode is described using a Gaussian beams model \cite{Janitz2015}, with a beam in diamond (blue) and air (orange), that are coupled at the diamond-air interface, where the beam widths match and the beam curvatures satisfy $n_dR_a=R_d$ for a planar diamond surface. The beam curvature of the air beam at the dimple follows the dimple's radius of curvature (ROC, here $25~\mu$m). The beam waist of the diamond beam ($w_{0,d}$) is fixed at the plane mirror, whereas the position of the air beam waist ($w_{0,a}$) at $z=\Delta z_a$ is obtained as a solution to the model.  
(b) Schematic of the cavity geometry. The dimple has a Gaussian shape with diameter $D_d$ (full width at $1/e$ of the Gaussian) and radius of curvature $ROC$, resulting in a minimum distance from fiber to mirror of $z_d$. The extent of the fiber ($D_f$) in combination with a fiber tilt $\theta$ result in an minimum extra cavity length of $z_f$. Figure is not to scale.
(c,d) Numerical (solid lines) and analytical (dashed lines) solutions for (c) $w_{0,d}$ and (d) the corresponding factor $g_0$ (\cref{eq:V_mod}) exhibit a stronger dependence on the air gap than on the diamond thickness, as described by \cref{eq:L_mod}. The exact analytic solution overlaps with the numerically obtained result.
(e,f) The ratio of the beam width on the concave mirror $w_m$ (e) and the dimple diameter $D_d$ determine the strength  of the clipping losses per round-trip (f). We here fix $t_d=4~\mu$m.   }
\label{fig:dimple_parameters}
\end{figure}

The laser-ablated dimple has a profile that is approximately Gaussian (\cref{fig:dimple_parameters}(b)). Beyond the radius ${D_d}/2$ the dimple significantly deviates from a spherical shape. If the beam width on the mirror ($w_m$) approaches this value, significant clipping losses result \cite{Hunger2010}:
	\begin{equation}
	\mathcal{L}_{clip}=\exp\left(-2\left(\frac{D_d/2}{w_m}\right)^2\right).
	\end{equation}
	 Using our coupled Gaussian beam model we find a numerical (\cref{fig:dimple_parameters}(e), solid line) and analytical (dashed line) solution to the beam width on the mirror and the resulting clipping losses (\cref{fig:dimple_parameters}(f)). 
	Like $w_{0,d}$, $w_m$ is influenced more strongly by the air gap width than by the diamond thickness. Consequently, the clipping losses are small even when the diamond membrane is relatively thick. For a Gaussian dimple with $ROC=25~\mu$m and $z_d = 0.3~\mu$m, we expect that $D_d\approx7.7~\mu$m. In this case for $t_d\approx 4~\mu$m and $t_a<2~\mu$m, the influence of clipping losses is negligible compared to other losses. 	
		The influence of clipping losses can be larger for cavity lengths at which transverse mode mixing appears \cite{Benedikter2015}. 
			
		Finally we note that the clipping losses should be treated in line with the method developed in \cref{sec:longitudinalconfinement}. The effective clipping losses are the clipping losses as found above, weighted by the relative field intensity in air (\cref{eq:rel_intensity}).

\section{Including real-world imperfections} \label{sec:optimal_finesse}

From the perspective of Purcell enhancement alone the requirements for the mirrors of our Fabry-Perot cavity are clear: since the Purcell factor is proportional to the quality factor of the cavity, high reflectivity of the cavity mirrors will provide the largest Purcell factor.

But when including real-world imperfections, we have to revisit this conclusion. In an open cavity system, having high-reflectivity mirrors comes with a price: the resulting narrow-linewidth cavity is sensitive to vibrations. And next to that, unwanted losses in the cavity force motivate an increase of the transmission of the outcoupling mirror, to detect the ZPL photons efficiently. In this section we analyse how both these effects influence the optimal mirror parameters.

\subsection{Vibration sensitivity} \label{sec:vibrationsensitivity}
The benefit of tunability of an open Fabry-Perot cavity has a related disadvantage: the cavity length is sensitive to vibrations. This issue is especially relevant for systems as considered here that require operation at cryogenic temperatures. Closed-cycle cryostats allow for stable long-term operation, but also induce extra vibrations from their pulse-tube operation. In setups specifically designed to mitigate vibrations passively \cite{Bogdanovic2017} vibrations modulate the cavity length over a range with a standard deviation of approximately $0.1$ nm. Here we discuss how to make a cavity perform optimally in the presence of such vibrations.

If vibrations change the cavity length, the cavity resonance frequency is modulated around the NV center emission frequency. For a bare cavity (with $\nu_\text{res} = mc/(2nL)$) the resonance frequency shift $d\nu_\text{res}$ due to vibrations over a characteristic (small) length $dL$ can be described by: 
\begin{equation} \label{eq:dnu_dL}
|d\nu_\text{res}|=\nu_\text{res} ~dL/L.
\end{equation}
Comparing this to the cavity linewidth $\delta \nu = \nu_\text{FSR}/F = c/{2nL F}$ and using $\nu_\text{res} = c/(n\lambda_\text{0,res})$ we find:
\begin{equation}
\frac{d\nu_\text{res}}{\delta\nu} = 2 \frac{dL}{\lambda_{0,res}} F.
\end{equation} 
For the impact of the vibrations the cavity length is thus irrelevant: rather the finesse plays an important role. If we demand that $d\nu_{res} < \delta\nu$ we find that we would need to limit the finesse to $F<\lambda_{0,res}/(2dL)$. 

For a hybrid cavity the frequency response is modified compared to the bare cavity situation by the influence of diamond-like and air-like modes. To find the modified response we evaluate the derivative of the resonance condition \cite{suppl} at the diamond-like and air-like mode:
\begin{equation} \label{eq:dnudta_hybrid}
\frac{d\nu_{a,d}}{dt_a}=-\frac{c}{(t_a+n_dt_d)\lambda_{0,res}}\left(1\pm\frac{n_d-1}{n_d+1}\frac{2n_dt_d}{t_a+n_dt_d}\right).
\end{equation}
The plus-sign on the left hand side corresponds to the case for an air-like mode, and the minus-sign corresponds to a diamond-like mode. A diamond-like mode is therefore less sensitive to vibrations than an air-like mode. This difference can be significant. For $t_d\approx4~\mu$m and $t_a\approx2~\mu$m, $\frac{d\nu_{a,d}}{dt_a}\approx 7~\text{GHz} /\AA$ in the air-like mode, while $\frac{d\nu_{a,d}}{dt_a}\approx 1~\text{GHz} /\AA$ in the diamond-like mode. The vibration susceptibility of a cavity with an AR coated diamond reduces to the bare cavity expression \cref{eq:dnu_dL}, with $L=t_a+n_dt_d+\lambda_{0}/2$, and thus takes an intermediate value between those for the air-like and diamond-like modes.

\begin{figure}[H]%
\centering
\includegraphics[width=0.9\columnwidth]{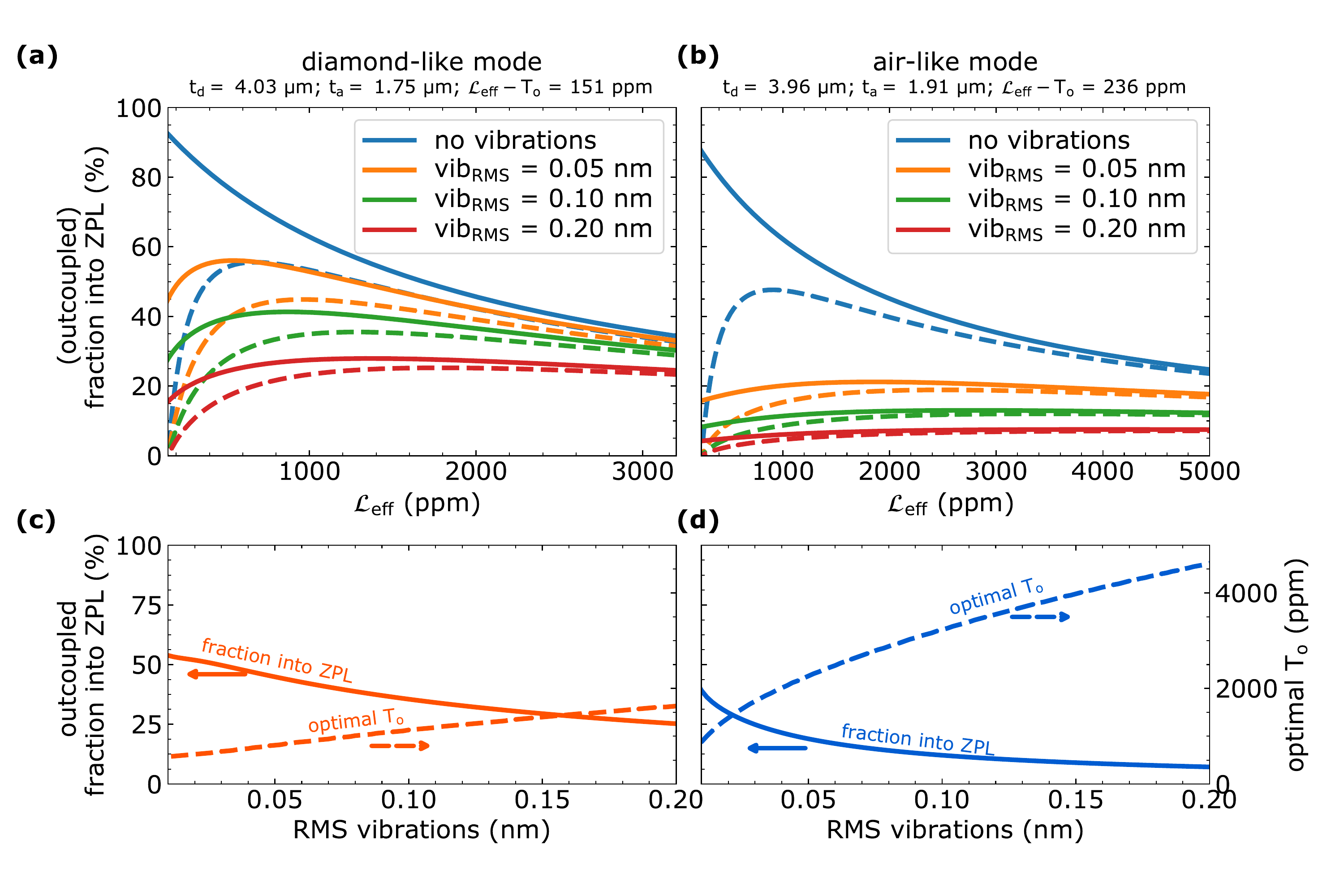}%
\caption{{\bf Optimal mirror parameters for a cavity under realistic conditions.} (a,b) Vibrations impact the average emission into the ZPL (solid lines) for (a) the diamond-like mode and more strongly for (b) the air-like mode. A reduced vibration sensitivity can be achieved for both by increasing the total cavity losses at the expense of a lower on-resonance Purcell factor. 
The fraction of ZPL photons outcoupled through the desired mirror (dashed line) can be increased by increasing the total losses via the transmittivity of the outcoupling mirror $T_\text{o}$. Outcoupling is assigned to be via the flat mirror, and the used parameters are $\LLMa=84$ ppm, $\LLMd = T_\text{o} + 34$ ppm, $\sDA=0.25$ nm RMS, and $ROC=20~\mu$m.
(c,d) By choosing an optimal $T_\text{o}$ (dashed line, right $x$-axis) the maximum outcoupled fraction into the ZPL for each level of vibrations (solid line, left $x$-axis) is obtained for (c) the diamond-like mode and (d) the air-like mode.}
\label{fig:intoZPLvslosses}%
\end{figure}

We include these vibrations in our model that describes the emission into the ZPL \cite{suppl}.  The results are shown as solid lines in \cref{fig:intoZPLvslosses}(a) and (b), for the diamond-like and air-like mode respectively. For a system with vibrations $\svib = 0.1$ nm, the emission into the ZPL for the diamond-like mode is $\approx 40\%$ for total losses of $\approx 800$ ppm, corresponding to a finesse of $F\approx8000$. 

The optimal losses may thus be higher than the minimal value set by unwanted losses. The losses can be increased by increasing the transmission through the outcoupling mirror. In this way not only vibration stability but also an improved outcoupling efficiency is achieved, as we see below.

\subsection{Outcoupling efficiency} \label{sec:outcouplingefficiency}
We do not only want to enhance the probability to emit a ZPL photon per excitation, but also want to couple this photon out of the cavity into the desired direction. The outcoupling efficiency is given by $\eta_\text{o} = T_\text{o}/\LLeff$, with $T_\text{o}$ the transmittivity of the outcoupling mirror. 
We choose to assign the plane mirror on the diamond side of the cavity as the outcoupling mirror. This assignment is motivated by comparison of the mode-matching efficiencies between the cavity mode and the dimpled fiber, and between the cavity mode and the free space path. For the free space path in principle perfect overlap with the Gaussian mode can be achieved, while for the fiber side this is limited to $\approx50\%$ for a cavity with $ROC=20~\mu$m, $t_d=4~\mu$m, and $t_a=2~\mu$m \cite{Joyce1984,Hunger2010,suppl}. Moreover, in this regime the mode-matching efficiency can only be improved by increasing each of these parameters, thereby compromising Purcell enhancement \cite{suppl}. 
Since the plane mirror is interfaced with diamond, we note that in the transmission $T_\text{o}$ this diamond termination has to be taken into account. When using a DBR stack with a high refractive index final layer, $T_\text{o}$ is higher than the transmission of the same stack in air.

The larger the unwanted losses ($\LLeff-T_\text{o}$) in the cavity are, the higher the transmission through the output mirror has to be to achieve the same outcoupling efficiency. The contributing unwanted losses are  transmission through the non-outcoupling mirror, scattering and absorption in both mirrors, and scattering at the diamond-air interface. Using values of $\approx 50$ ppm, $\approx 24 $ ppm and $\approx 10$ ppm for mirror transmission, scattering and absorption \cite{Bogdanovic2017}, and a diamond-air interface roughness of $\sDA=0.25$ nm \cite{Appel2016,Bogdanovic2017b,Riedel2017a}, we find that the unwanted losses are 139 ppm (236 ppm) for the diamond-like (air-like) mode using the analytic expression from \cref{eq:L_M,eq:L_SAD}.

An outcoupling efficiency $\eta_0>0.5$ is then achieved for $T_\text{o}>139$ ppm (236 ppm). The additional losses this would add to the cavity system are less than what is optimal for typical vibrations of $\svib\approx0.1$ nm ($\rightarrow\LLeff\approx800$ ppm (3000 ppm)) for both the diamond-like and air-like modes. Vibrations thus have a dominant effect. To improve the cavity performance in this regime focus should thus be on the reduction of vibrations over the reduction of unwanted losses. A possible route for vibration reduction is by extending active cavity stabilisation techniques for Fabry-Perot cavities \cite{Khudaverdyan2008, Gallego2015, Brachmann2016,Janitz2017} to operation under pulse-tube conditions. 

Including the outcoupling efficiency in our model we find the fraction of photons that upon NV excitation are emitted into the ZPL and subsequently coupled out of the cavity into the preferred mode (dashed lines in \cref{fig:intoZPLvslosses}(a),(b)). For each value of vibrations, we can maximize this fraction by picking an optimal $T_\text{o}$.  For the diamond-like and air-like mode the results of this optimization are shown in \cref{fig:intoZPLvslosses}(c),(d). For vibrations of $0.1$ nm, the best results ($\approx 35\%$ probability of outcoupling a ZPL photon) are expected to be achieved in a diamond-like mode with $T_\text{o}\approx1200$ ppm. We note that this corresponds to a modest Purcell factor of 40, leading to an excited state lifetime reduction to 5.2 ns, and a lifetime-limited linewidth of 31 MHz. Purcell factors higher than this lead to increased linebroadening, which should be taken into account for optical excitation, see e.g. Ref \cite{Hanks2017}. Increased Purcell factors at such levels have a small effect on the resulting emission into the ZPL (\cref{eq:branching}), and thus a limited benefit for an optimal design.

\section{Conclusions} \label{sec:conclusions}
In summary, we have developed analytical descriptions giving the influence of key parameters on the performance of a Fabry-Perot cavity containing a diamond membrane. This analytical treatment allows us to clearly identify sometimes conflicting requirements and guide the optimal design choices.

We find that the effective losses in the cavity are strongly dependent on the precise diamond thickness. This thickness dictates the distribution of the electric field in the cavity, with as extreme cases the diamond-like and air-like modes in which the field lives mostly in diamond and air respectively. As a result, the losses due to the mirror on the air side are suppressed by a factor $n_d$ in diamond-like modes while they are increased by the same factor in the air-like modes. In contrast the losses resulting from diamond surface roughness are highest in the diamond-like mode. The two types of losses can therefore be traded-off against each other. If the diamond surface roughness can be made sufficiently low ($<0.4$ nm RMS for mirror losses on the air gap side of 85 ppm), the total losses are lowest in the diamond-like mode.

The transverse confinement of the cavity is captured in a geometrical factor $g_0$ that depends on the beam waist alone. It is determined by the radius of curvature of the dimple and an expression that captures the effect of the cavity component thicknesses: $t_a + t_d/n_d$. The width of the air gap $t_a$ thus has a dominant influence, while the influence of the diamond thickness $t_d$ is reduced by the diamond refractive index $n_d$. From a geometrical perspective, the focus in the cavity design should thus be on small radii of curvature and small air gaps. 

Although the highest Purcell factors are achieved for low cavity losses, vibrational instability of the cavity length and the presence of unwanted losses suggest that lowering the cavity finesse can be advantageous. We find that a cavity supporting an air-like mode is more severely affected by vibrations than one supporting a diamond-like mode. For example, for vibrations of $0.1$ nm RMS and unwanted losses of $\approx 190$ ppm we find that the optimal fraction of ZPL photons reaching the detector is obtained with a diamond-like mode and an outcoupling mirror transmission of $T_\text{o} \approx 1200$ ppm. 

The experimentally realistic parameter regimes considered here include a 4 $\mu$m diamond thickness to support optically coherent NV centres and vibrations of $0.1$ nm RMS under pulse-tube operation with passive stabilisation. In this regime with an optimized design an emission efficiency of ZPL photons into the desired outcoupled optical mode after resonant excitation of 35\% can be achieved. This constitutes a two orders of magnitude improvement compared to existing approaches, for which the branching ratio into the ZPL is $\approx3\%$ and the collection efficiencies are typically $\approx 10\%$ \cite{Hensen2015a}.

Purcell enhancement with open Fabry-Perot cavities will open the door to efficient spin-photon interfaces for diamond-based quantum networks. The analysis presented here clarifies the design criteria for these cavities. Future experimental design and investigation will determine how to combine such cavities with resonant excitation and detection for spin-state measurement \cite{Robledo2011} and long distance entanglement generation \cite{Hensen2015a,Bock2018,Dreau2018}.

\begin{acknowledgements}
We thank J. Benedikter, L. Childress, A. Galiullin, E. Janitz, S. Hermans, P.C. Humphreys, and D. Hunger for helpful discussions. We acknowledge support from the Netherlands Organisation for Scientific Research (NWO) through a VICI grant, the European Research Council through a Synergy Grant, and the Royal Netherlands Academy of Arts and Sciences and Ammodo through an Ammodo KNAW Award.
\end{acknowledgements}

\bibliography{Manuscript}

\begin{thebibliography}{49}
\bibitem [{cod()}]{code}%
  \BibitemOpen
  \href@noop {} {}\bibinfo {note} {We are happy to provide the code on
  request.}\BibitemShut {Stop}%
\end{thebibliography}

\end{document}